\documentclass[aps,prc,twocolumn,superscriptaddress,floatfix,showpacs,nofootinbib]{revtex4-2}
\usepackage{multirow}
\usepackage{graphicx}
\usepackage{color}
\usepackage{amsmath}
\usepackage{amssymb}
\usepackage[colorlinks=true]{hyperref}
\usepackage{tikz}
\usepackage{etoolbox}
\usepackage{bm}
\usepackage[utf8]{inputenc}
\usepackage{cleveref}
\usepackage{amsmath}
\usepackage{amssymb}
\usepackage{verbatim}
\usepackage{float}
\usepackage[bottom]{footmisc}

\newcommand {\trento}   {{\tt TRENTo}}

\newcommand {\urqmd}    {{\tt UrQMD}}

\newcommand {\vis}      {{\tt VISH2+1}}

\newcommand {\hcf}     {{\tt Hydro-Coal-Frag}}
\newcommand {\hf}     {{\tt Hydro-Frag}}
\newcommand {\pythia}   {{\tt PYTHIA8}}
\newcommand {\lbt}   {{\tt LBT}}
\newcommand {\kln}   {{\tt KNL}}
\newcommand {\ekrt}   {{\tt EKRT}}
\newcommand {\wn}   {{\tt WN}}

\newcommand{\n}{\nonumber}

\definecolor{dgreen}{cmyk}{1.,0.,1.,0.4}        
\definecolor{orange}{cmyk}{0.,0.353,1.,0.}    

\begin{document}
\title{
Exploring the partonic collectivity in small systems at the LHC}

\author{Yuanyuan Wang}
\affiliation{School of Physics, Peking University, Beijing 100871, China}

\author{Wenbin Zhao}
\email{Wenbin Zhao: WenbinZhao@lbl.gov}
\affiliation{Nuclear Science Division, Lawrence Berkeley National Laboratory, Berkeley, California 94720, USA}
\affiliation{Physics Department, University of California, Berkeley, California 94720, USA}

\author{Huichao Song}
\email{Huichao Song: huichaosong@pku.edu.cn}
\affiliation{School of Physics, Peking University, Beijing 100871, China}
\affiliation{Center for High Energy Physics, Peking University, Beijing 100871, China}
\date{\today}
\begin{abstract}
Using the \hcf{} model that combines hydrodynamics at low $p_{\rm T}$, quark coalescence at intermediate $p_{\rm T}$, and the \lbt{} transport model at high $p_{\rm T}$, we study the spectra and elliptic flow of identified hadrons in high multiplicity p--Pb and p--p collisions at the Large Hadron Collider (LHC). In p--Pb collisions, the \hcf{} model gives a good description of the differential elliptic flow over the $p_{\rm T}$ range from 0 to 6 GeV and the approximate number of constituent quark (NCQ) scaling at intermediate $p_{\rm T}$.  Although \hcf{} model can also roughly describe the elliptic flow in high multiplicity p--p collisions with the quark coalescence process, the larger contribution from the string fragmentations leads to a notable violation of the NCQ scaling of $v_2$ at intermediate $p_{\rm T}$ as observed in the experiment. Comparison runs of the \hf{} model without the coalescence process demonstrate that regardless the parameter adjustments, the \hf{} model cannot simultaneously describe the $p_{\rm T}$ spectra and the elliptic flow of identified hadrons in either p--Pb collisions or p--p collisions. The calculations in this paper thus provide support for the existence of partonic degrees of freedom and the possible formation of the QGP in the small systems created at the LHC.
\end{abstract}



\maketitle

\clearpage

\section{Introduction}
\label{section1}
The main goals of the heavy ion program at the Relativistic Heavy Ion Collider (RHIC), and the Large Hadron Collider (LHC) are to create and study the quark--gluon plasma (QGP), a state of hot and dense nuclear matter in which quarks and gluons are no longer bound into hadrons. Since the running of 200 A GeV Au--Au collisions at RHIC, many signals of the QGP have been observed, such as the collective flow, valence quark scaling and jet quenching~\cite{Gyulassy:2004vg, STAR:2005gfr, Muller:2006ee, 796947, Jacobs:2007dw}. It was found that the QGP droplet created in relativistic heavy ion collisions is the most perfect liquid in nature, with collective features well described by the relativistic hydrodynamics~\cite{Heinz:2013th, Gale:2013da, Shuryak:2014zxa,Song:2013gia,Song:2017wtw}.
 
For small systems, the multi-particle correlations in high multiplicity p--Pb, and p--p collisions at the LHC have exhibited surprising collective behaviors, including the long--range double--range structures in two--particle correlations~\cite{CMS:2015fgy, ATLAS:2015hzw,  ATLAS:2013jmi, CMS:2015yux, ALICE:2019zfl, CMS:2010ifv, CMS:2012qk, ALICE:2012eyl}, four--particle cumulant $c_2\{4\}$ turning to negative values in high multiplicity events~\cite{CMS:2016fnw, ATLAS:2013jmi, CMS:2015yux, ALICE:2019zfl, ALICE:2014dwt}, and mass splitting of elliptical flow among different hadron species~\cite{CMS:2016fnw, ALICE:2013snk, CMS:2014und}. In the past ten years, various
models have been applied to explain the emergence of collectivity in the small systems. Hydrodynamic models based on final state interactions~\cite{Bozek:2011if, Nagle:2013lja, Schenke:2014zha,  Shen:2016zpp, Zhao:2017rgg, Mantysaari:2017cni, Werner:2013tya,  Qin:2013bha, Zhou:2015iba, Bozek:2015swa, Weller:2017tsr, Zhao:2020pty, Bzdak:2014dia, Kurkela:2018ygx, Zhao:2022ugy, Zhao:2022ayk, Wu:2023vqj} and the Color Glass Condensate (CGC) model based on initial state correlations~\cite{Dusling:2012iga, Muneyuki:2012ur, Dusling:2012cg, Kovner:2012jm, Kovchegov:2012nd, Schenke:2015aqa, Lappi:2015vta, Schenke:2016lrs, Dusling:2017dqg, Dusling:2017aot, Mace:2018yvl, Schenke:2021mxx} can both qualitatively describe the collective flow in the low transverse momentum region. In the high transverse momentum region, jet quenching, the hard probe signals for the QGP formation, has not been observed due to the limited size and lifetime of the small systems. The experimentally observed hadron suppression factor $R_{\rm{pA}}$ of the small system is consistent with the calculations from the cold nuclear effect~\cite{Albacete:2016veq,  Eskola:2016oht}. 

In the intermediate transverse momentum regime, the ALICE and CMS collaborations have observed an approximate number of constituent quark (NCQ) scaling of $v_2$ for identified hadrons in high multiplicity p--Pb collisions at the LHC~\cite{Pacik:2018gix, CMS:2018loe}. Meanwhile, the \hcf{} model that includes various hadronization mechanisms from hydrodynamics, quark coalescence, and the \lbt{} string fragmentations in different transverse momentum regimes has been developed. It has well described the transverse momentum spectrum and differential elliptic flow for identified hadron from 0 to 6 GeV in p--Pb and Pb--Pb collisions~\cite{Zhao:2020wcd, Zhao:2020wpi, Zhao:2021vmu} and explained the NCQ scaling of $v_2$ in p--Pb collisions with the quark coalescence process. 

Recently, the CMS, ATLAS and ALICE collaborations have measured the $p_{\rm T}$--spectra and  $v_2(p_{\rm T})$ of all charged and identified hadrons in high multiplicity p--p collisions at $\sqrt{s_{\rm{NN}}}=13$ TeV~\cite{ALICE:2020nkc, CMS:2016fnw, ATLAS:2016yzd}.  In this paper, we will perform the first systematic study of the elliptic flow over the transverse momentum region from 0 to 6 GeV in both high multiplicity p--p and p--Pb collisions.  We will show that, with the quark coalescence process, the \hcf{} model  describes the elliptic flow of various hadron species in both  p--p and p--Pb collisions, which also gives an approximate NCQ scaling of $v_2$ at intermediate $p_{\rm T}$ in p--Pb collisions. While the larger contribution from the string fragmentations leads to a notable violation of the NCQ scaling of $v_2$ in p--p collisions as observed in the experiment. Compared to our early short paper~\cite{Zhao:2020wcd}, we will also explain more details about the \hcf{} model for the sake of documentation and easiness of reading.

\section{The model and set--ups}
\label{sec:model}

Relativistic heavy ion collision is a dynamical evolution process, which is consistent with the initial state, pre-equilibrium dynamics, thermalization, the QGP expansion, hadronization, hadronic evolution, and the chemical and kinematic freeze--out. In this work, we implement the \hcf{} model, which produces low $p_{\rm T}$ hadrons from hydrodynamics, high $p_{\rm T}$ hadrons from fragmentations and intermediate $p_{\rm T}$ hadrons from the coalescence process~\cite{Zhao:2020wcd}. Here, the initial entropy distributions are generated by a parameterized initial condition model
 \trento{} (Reduced Thickness Event--by--Event Nuclear Topology)~\cite{Moreland:2014oya}, and the evolution of the QGP is described by a (2+1)--dimensional viscous hydrodynamics (\vis{})~\cite{Heinz:2005bw, Song:2007fn, Song:2007ux, Song:2009gc, Shen:2014vra}, which produces thermalized hadrons and partons at low $p_{\rm T}$ from the freeze--out surface near $T_{\rm c}$. The initial hard partons are produced by \pythia{}~\cite{Sjostrand:2007gs}, and the interaction between the initial hard partons and the bulk thermal medium is modeled by the linear Boltzmann transport model (\lbt{})~\cite{Wang:2013cia, Peng:2014jia, Cao:2016gvr, Xing:2019xae}, which includes both pQCD elastic scattering and medium--induced gluon radiation within the high--twist approach. Hadron production at intermediate $p_{\rm T}$ is described by the parton coalescence model, which includes thermal--thermal, thermal--hard, and hard--hard partons recombinations with the remaining hard partons hadronized by string fragmentation~\cite{Zhao:2020wcd, Zhao:2020wpi}.  Finally, these produced hadrons at different $p_{\rm T}$ regimes are fed to the Ultra-relativistic Quantum Molecular Dynamics (\urqmd{}) for subsequent hadronic evolution with scatterings and resonance decays~\cite{Bass:1998ca}. 

More specifically, for soft hadron production at low $p_{\rm T}$, we implement a 2+1D viscous fluid model, \vis, with longitudinal boost--invariance to simulate the collective expansion of the QGP, which numerically solves the transport equations of the energy--momentum tensor $T^{\mu\nu}$ and the evolution equations of the shear viscous tensor $\pi^{\mu\nu}$ and the bulk viscous pressure $\Pi$ in the second--order~\cite{Song:2007fn, Song:2007ux, Song:2008si, Song:2009gc}. To close the systems, we input the equation of state s95--PCE~\cite{Huovinen:2009yb, Shen:2010uy}, and start the hydrodynamic simulations at $\tau=0.8$ fm/c with the \trento{} model, which parameterizes the initial entropy density by the thickness functions $T_{\rm A/B}$: $s=s_0[(T^p_{\rm A}+T^p_{\rm B})/2]^{1/p}$. Here, $p$ is an adjustable parameter that allows \trento{} to effectively transit between different initial conditions, such as \kln, \ekrt, \wn, etc.~\cite{Moreland:2014oya,Bernhard:2016tnd, Moreland:2014oya}. In this paper, we implement the version of \trento{} with sub-nucleonic structures~\cite{Bernhard:2016tnd, Moreland:2018gsh}. So that the thickness function is written as $T(x, y)\equiv \int {\rm d}z\, \frac{1}{n_{\rm c}} \sum_{i=1}^{n_{\rm c}} \gamma_i\, \rho_{\rm c} \,({\bm x} - {\bm x}_i \pm {\bm b}/2)$, where $n_{\rm c}$ is the number of constituents in a nucleon, $\rho_{\rm c}$ is the density of the constituent with a Gaussian form: $\rho_{\rm c}={(2\pi w_{\rm c}^2)^{-3/2}} \exp{[-{\bm x^2}/({2 w_{\rm c}^2})]}$, $ \gamma_i$ is a random weighting factor with a unit mean and variance $1/\sigma_{\rm flut}$,  ${\bm x}_i$ is the position of the constituent and ${\bm b}$ is the impact parameter. As the QGP fireball expands and cools down, the hydrodynamic system hadronizes near $T_{\rm c}$, which is realized by the iSS event generator~\cite{Song:2010aq} that produces thermal hadrons on the freeze--out surface according to the Cooper--Frye formula~\cite{Song:2010aq}. For the following quark coalescence process at intermediate $p_{\rm T}$, we also sample the thermal partons at low $p_{\rm T}$ from the hydrodynamic freeze-out surface near $T_{\rm c}$. 

Following the early model calculations~\cite{Li:2019nzj, Carroll:2008sv, Gao:2017gvf, Fu:2019hdw, Levai:1997yx, Zhao:2020wcd, Zhao:2020wpi}, we set the mass of the thermal quark to $m_{{\rm u}, {\rm d}}=0.25$ GeV, $m_{\rm s}=0.43$ GeV, and convert the thermal gluons into to quark and anti-quark pairs via $gg \rightarrow q \bar{q}$.  For for p--p collisions at $\sqrt{s_{\rm{NN}}}=13$ TeV, the \trento{} parameters are set to: the normalization factor $s_0=20$, the reduced thickness $p=0$, the fluctuation $\sigma_{\rm flut}=0.19$, the nucleon width  $r_{\rm cp}=0.92$ fm, the constituent width $w_{\rm c}=0.6$ fm, and the constituent number $n_{\rm c}=6$~\cite{Bernhard:2016tnd}. We also tune the specific shear and bulk viscosity and the hydrodynamic switching temperature to: $\eta/s=0.03$, $\zeta/s=0.4$ and $T_{\rm {switch}}=150$ MeV in order to fit the soft $p_{\rm T}$--spectra and elliptic flow blow 2 GeV. For p--Pb collisions at $\sqrt{s_{\rm{NN}}}=5.02$ TeV, the \trento{} parameters are the same as the p--p collisions except that the constituent width is changed to $w_{\rm c}=0.4$ fm, and the switching temperature is set to: $T_{\rm {switch}}=160$ MeV. The specific shear and bulk viscosity are tuned as: $\eta/s=0.12$ and $\zeta/s=0.6$ to fit the new ALICE flow data in p--Pb collisions obtained by the ``template fit'' method~\cite{ALICE:2023mingrui}. Note that, due to the different non-flow subtraction assumptions, the ``template fit'' results generally have larger flow magnitudes than those obtained from the ``peripheral subtraction'' method used by the CMS collabration~\cite{CMS:2018loe}.

For hard parton production, we implement \pythia~\cite{Sjostrand:2007gs} to produce the initial hard partons, and then feed them into the \lbt{} model to simulate their propagation in the bulk medium.  The \lbt{} model solves the linear Boltzmann equations for quarks, antiquarks, and gluons with elastic scatterings,  inelastic scatterings and medium--induced gluon radiation within an expanding QGP described by hydrodynamics~\cite{Wang:2013cia, Peng:2014jia, Cao:2016gvr,  Chen:2017zte, Luo:2018pto, Xing:2019xae}. For p--Pb collisions, the nuclear shadowing effects on the momentum space distribution of the initial hard partons inside the lead have been parameterized by {\tt EPPS16} in \pythia~\cite{Eskola:2016oht}. The only tunable parameter in \lbt{} is the strong coupling constant $\alpha_{\rm s}$, which is set to $\alpha_{\rm s}=0.15$ that has been tuned to fit the $R_{\rm AA}$ and the anisotropic flow of light and heavy flavors at high $p_{\rm T}$ for Au--Au collisions and Pb--Pb collisions at RHIC and the LHC~\cite{He:2015pra, Cao:2017hhk}. The virtualities of hard gluons are generally set to be zero in {\tt LBT}.  Following~\cite{Han:2016uhh},  we assume that the hard gluon has a virtual mass uniformly distributed between $2m_{\rm s}$ and $m_{\rm{max}}$, where $m_{\rm s}$ is the mass of the strange quark, and $m_{\rm{max}}$ is a tunable parameter that determines the ratio of light to strange quarks for the following quark coalescence calculations. According to Ref.~\cite{Zhao:2020wcd}, we set $m_{ \rm {max}}=1.5$ GeV to fit the relative spectra between kaon and pion at intermediate $p_{\rm T}$ in our model calculations. With virtual masses, the hard gluons decay isotropically into $q\bar{q}$ pairs, with  $u\bar{u}$ and $d\bar{d}$ pairs had equal decay weight, and the ratio of light to strange quarks given by the associated phase space. These final partons are then either fed into the following coalescence model calculations or form hard hadrons through string fragmentations.

For hadron production at intermediate $p_{\rm T}$, we implement the quark coalescence model, which combines thermal--thermal, thermal--hard, and hard--hard partons obtained from the hydrodynamic freeze-out and the LBT parton shower evolution. Using the thermal and hard parton distributions, obtained from the \vis{} and \lbt{} models, the momentum distributions of mesons and baryons from coalescence  are generated from an overlap of the Wigner functions~\cite{Han:2016uhh}:
\begin{align}
\frac{{\rm d}N_{\rm M}}{{\rm d}^3{\bm{P}}_{\rm M}}&=g_{\rm{M}}\int{\rm{d}}^3{\bm{x}}_1 {\rm{d}}^3{\bm{p}}_1 {\rm{d}}^3{\bm{x}}_2 {\rm{d}}^3{\bm{p}}_2     f_{\rm q}({\bm{x}}_1, {\bm{p}}_1) f_{\bar{\rm q}}({\bm{x}}_2,{\bm{p}}_2) \n\\
&\times W_{\rm M}({\bm y}, {\bm k})\delta^{(3)}({\bm P}_{\rm M}-{\bm p}_1-{\bm p}_2), \n\\
\frac{{\rm d}N_{\rm B}}{{\rm d}^3{\bm P}_{\rm B}}&=g_{\rm{B}}\int{\rm d}^3{\bm x}_1 {\rm d}^3{\bm p}_1 {\rm d}^3{\bm x}_2 {\rm d}^3{\bm p}_2
{\rm d}^3{\bm x}_3 {\rm d}^3{\bm p}_3 \n\\
&\times f_{{\rm q}_1}({\bm x}_1,{\bm p}_1) 
f_{{\rm q}_2}({\bm x}_2, {\bm p}_2) 
f_{{\rm q}_3}({\bm x}_3,{\bm p}_3) \n\\
&\times W_{\rm B}({\bm y}_1,{\bm k}_1;{\bm y}_2,{\bm k}_2)\delta^{(3)}({\bm P}_{\rm B}-{\bm p}_1-{\bm p}_2-{\bm p}_3). \n
\end{align}
where $f_{\rm q}$ and $f_{\bar{\rm q}}$ are the phase--space distributions of quarks and anti-quarks, $W_{\rm M}$ and $W_{\rm B}$ are the Wigner functions of the meson and the baryon, $\bm y$ and $\bm k$ are the relative coordinates and relative momenta between the valence quarks in the local rest frame of the meson. According to Ref.~\cite{Han:2016uhh}, the Wigner function of meson in the $n$--th excited state is $W_{{\rm M}, n}({\bm y}, {\bm k})=v^n e^{-v}/{n!}$, where $v=({\bm y}^2/{\sigma^2_{\rm M}}+{\bm k}^2{\sigma^2_{\rm M}})/2$. The Wigner function of a baryon in the $n_1$--th and $n_2$--th excited states is given by $W_{{\rm B}, n_1, n_2} ({\bm y}_1, {\bm k}_1; {\bm y}_2, {\bm k}_2)={{v}_1^{n_1}}e^{-{v}_1}{{v}_2^{n_2}}e^{-{v}_2}/({n_1!}{n_2!})$, where ${v}_i=({{\bm y}_i ^2}/{\sigma_{{\rm B}_i}^2}+{\bm k}_{i} ^2 \sigma_{{\rm B}_i}^2)/2$  with ${\bm y}_i$ and ${\bm k}_i$ being the relative coordinates and momenta among the three constituent quarks in the local rest frame of the baryon. Following Refs.~\cite{Zhao:2020wcd, Zhao:2021vmu}, we take the excited meson states up to $n=10$ and the excited baryon states up to $n_1+n_2=10$~\cite{Han:2016uhh}. The width parameters $\sigma_{\rm M}$, $\sigma_{\rm {\rm B}_1}$ and $\sigma_{\rm {\rm B}_2}$ are determined by the radii of formed hadrons from the Particle Data Group~\cite{ParticleDataGroup:2012pjm}. After the thermal--thermal, thermal--hard, and hard--hard partons have recombined, the remaining hard partons that do not have coalescence partners are connected with strings, which then perform the string fragmentation of the hardon standalone mode in \pythia{} to form hard hadrons~\cite{Sjostrand:2007gs}.  For more details, please refer to \cite{Han:2016uhh, Zhao:2020wcd}.

In short, the \hcf{} model combines hydrodynamics to produce low $p_{\rm T}$ hadrons, quark coalescence model to produce intermediate $p_{\rm T}$ hadrons, and \lbt{} string fragmentations to produce high $p_{\rm T}$ hadrons. After the system has completed all the hadronization processes, the subsequent hadronic evolution is simulated by \urqmd{}~\cite{Bass:1998ca}, which propagates hadrons with elastic and inelastic scatterings and resonance decays until the system becomes very dilute with a kinetic freeze--out. To avoid double counting within \hcf{}, we set a $p_{\rm T}$ cut--off between hadrons produced by hydrodynamics and those produced by quark coalescence. We only recombine these thermal partons with transverse momentum $p_{\rm T} \textgreater p_{\rm {T_1}}$ and produce thermal mesons and thermal baryons below $2p_{\rm {T_1}}$ and $3p_{\rm {T_1}}$  with hydrodynamics. Meanwhile, for the \lbt{} evolution and fragmentations, we count only these final hard hadrons with transverse momentum  $p_{\rm T} \textgreater p_{\rm {T_2}}$ to avoid overcounting with these low $p_{\rm T}$ hadrons produced from hydrodynamics. The two parameters $p_{\rm {T_1}}$ and $p_{\rm {T_2}}$ and the above parameter $m_{\rm{max}}$ for gluon virtuality are determined by fitting the transverse momentum spectrum of $\pi$, K, and P and the ${\rm P}/\pi$ ratio at the transition range $p_{\rm T}\sim 2-4$ GeV  in the high multiplicity events. Here we find  $p_{\rm {T_1}}=1.4$ GeV and $p_{\rm {T_2}}=2.0$ GeV for p--p collisions and $p_{\rm {T_1}}=1.6$ GeV and $p_{\rm {T_2}}=2.6$ GeV for p--Pb collisions, together with $m_{\rm{max}}=1.5$ GeV, can well describe these measured $p_{\rm T}$ spectra in the high multiplicity.  With the fixed parameters described above, we further predict the differential elliptic flow of $\pi$, K, and P and the associated NCQ scaling for $105 \textless N_{\rm{ch}} \textless 150$  for p--p collisions at $\sqrt{s_{\rm{NN}}}=13$ TeV and for $0-20\%$ centrality p--Pb collisions at $\sqrt{s_{\rm{NN}}}=5.02$ TeV. For comparison, we also perform the calculations from the \hf{} model, which includes only hydrodynamics and string fragmentations but without the quark coalescence process. 

\begin{figure*}[thb]
\begin{center}
\includegraphics[width=0.8\textwidth]{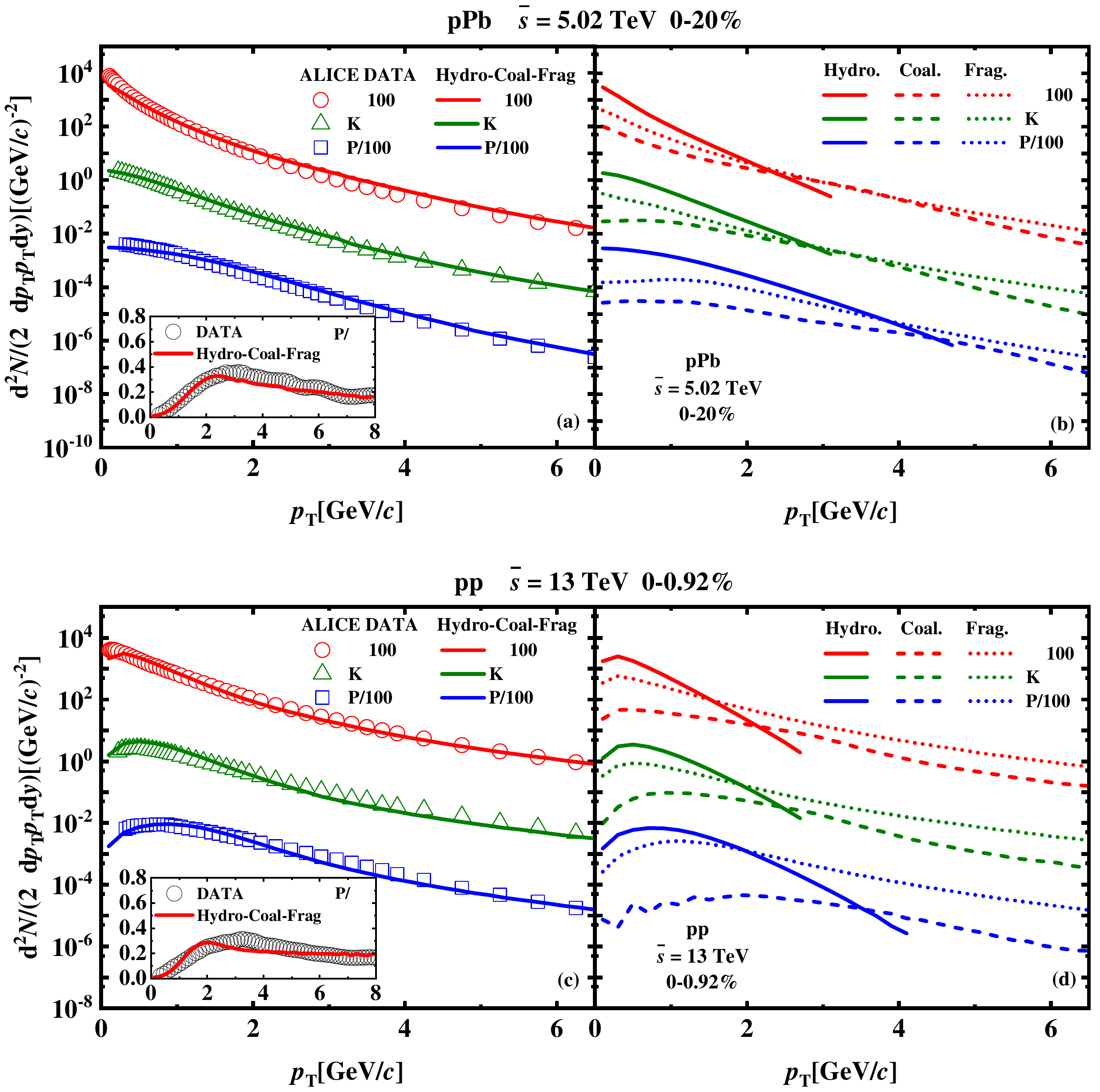}
\caption{(Color online) Left panel: transverse momentum spectra of $\pi$, K and P calculated from the \hcf{} model, with the inset panels showing the proton--to--pion ratio. Right panel: the contributions from hydrodynamics (solid lines), quark coalescence (dashed lines), and string fragmentation (dotted lines) processes. Upper and bottom panels are for 0-20\%  p--Pb collisions at $\sqrt{s_{\rm{NN}}}=5.02$ TeV and for {$0-0.92\%$ p--p collisions} at $\sqrt{s_{\rm{NN}}}=13$ TeV , respectively. The data are from the ALICE papers~\cite{ALICE:2020nkc, ALICE:2016dei}.}
\label{f:spectraall}
\end{center}
\end{figure*}

\begin{figure*}[thb]
\begin{center}
\includegraphics[width=0.8\textwidth]{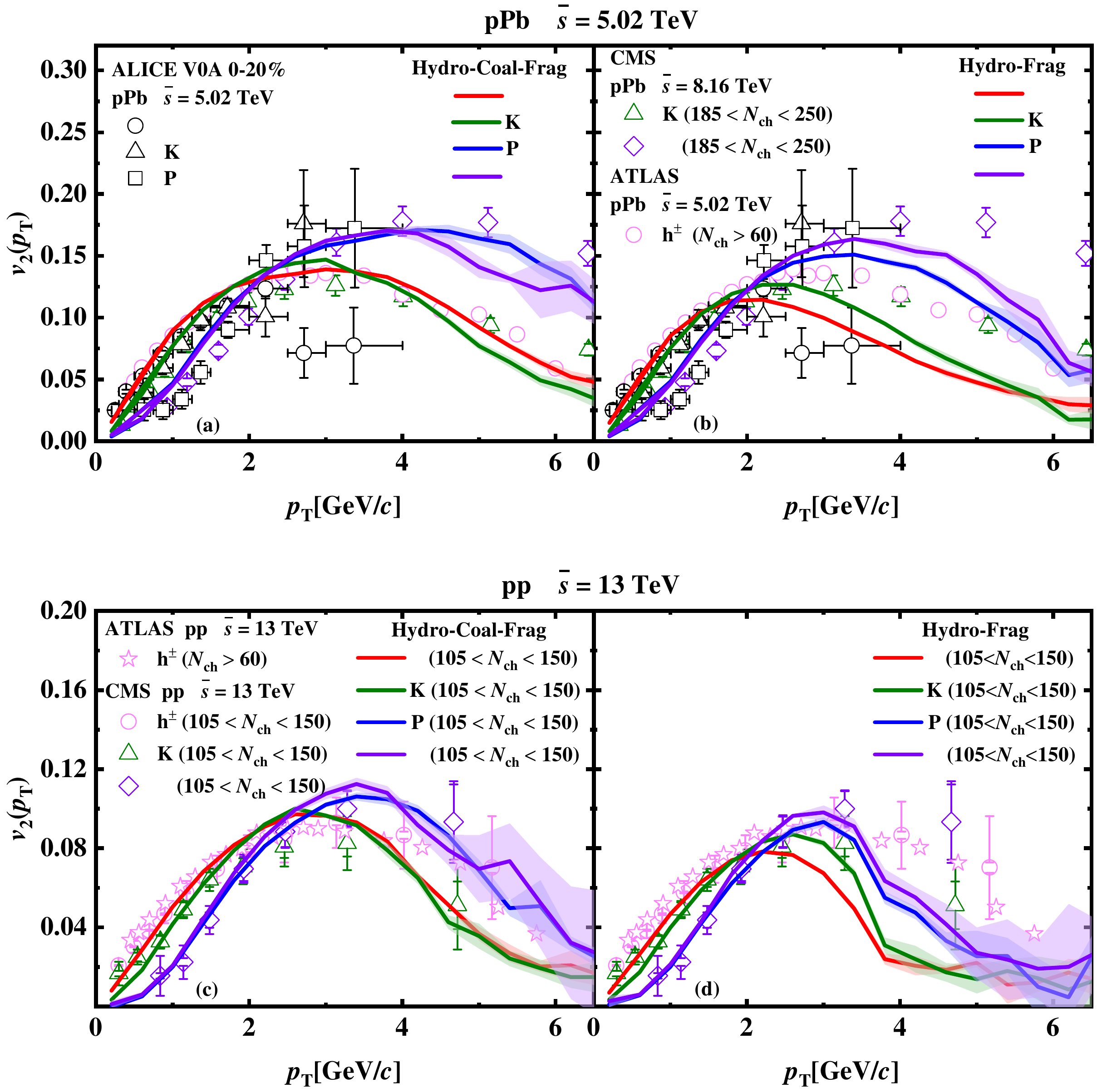}
\caption{(Color online) The differential elliptic flow of $\pi$, K, P and $\Lambda$  in 0-20\%  p--Pb collisions at $\sqrt{s_{\rm{NN}}}=5.02$ TeV (upper panel) and in $105 \textless N_{\rm{ch}}\textless 150$ p--p collisions at $\sqrt{s_{\rm{NN}}}=13$ TeV (bottom panel), calculated from the \hcf{} model and the \hf{} model, respectively. Data are from the CMS, ATLAS, and ALICE papers~\cite{CMS:2016fnw, ATLAS:2016yzd, CMS:2018loe, ALICE:2013snk}. }
\label{f:v2all}
\end{center}
\end{figure*}

\begin{figure*}[thb]
\begin{center}
\includegraphics[width=0.8\textwidth]{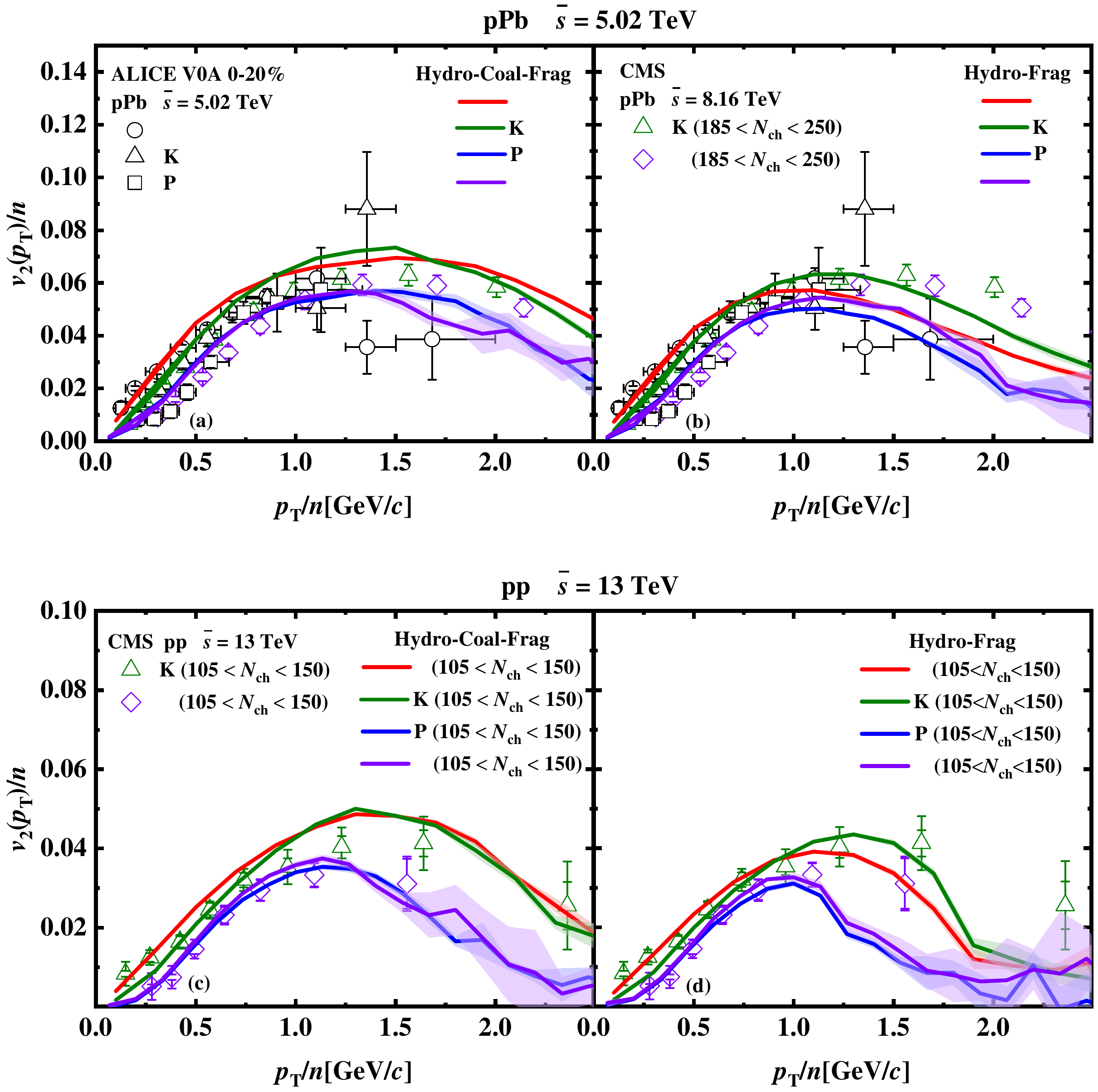}
\caption{(Color online) Similar to Fig.~\ref{f:v2all}, but for the number of constituent quark (NCQ) scaled $v_2(p_{\rm T})$ of $\pi$, K, P and $\Lambda$. }
\label{f:v2nall}
\end{center}
\end{figure*}

\section{Results and Discussions}

The left panels (a) and (c) of Fig.~\ref{f:spectraall}  show the transverse momentum spectra of $\pi$, K and P in $0-20\%$ p--Pb collisions at $\sqrt{s_{\rm{NN}}}=5.02$ TeV and in {$0-0.92\%$ p--p collisions} at $\sqrt{s_{\rm{NN}}}=13$ TeV, calculated from the \hcf{} model with the coalescence process. With properly tuned parameters, the model gives good descriptions of the transverse momentum spectra of $\pi$, K and P
over the  $p_{\rm T}$ range from 0 to 6 GeV. It also nicely describes the  ${\rm P}/\pi$ ratio as a function of $p_{\rm T}$, showing an increasing then a decreasing trend, as observed by the ALICE collaboration.
The two right panels (b) and (d) show the calculated $p_{\rm T}$ spectra from the \hcf{} model with the individual contributions from hydrodynamics, quark coalescence and string fragmentation. It demonstrates that hadrons produced at low or high transverse momentum are dominantly contributed by hydrodynamics or the string fragmentation respectively. At intermediate $p_{\rm T}$ regime, both quark coalescence and string fragmentation contribute to the hadron productions. It is noteworthy that the relative yield from quark coalescence is more pronounced in p--Pb collisions than in p--p collisions. Specifically, in p--Pb collisions, the yield of coalescence hadrons is close to that of fragmentation hadrons at intermediate $p_{\rm T}$, while in p--p collisions, coalescence hadrons is below 50\% of the fragmentation hadrons. This difference is due to the larger QGP droplet in p--Pb collisions, which generates more thermal partons and leads to a larger yield of coalescence hadrons than that in p--p collisions. With these different hadron production procedures that smoothly transit among different transverse momentum regimes, the \hcf{} model can well describe the $p_{\rm T}$ spectra of $\pi$, K and P from $0-6$ GeV in both p--Pb and p--p collisions. 

Figure.~\ref{f:v2all} shows the differential elliptic flow $v_2(p_{\rm T})$ of $\pi$, K, P and $\Lambda$ in  $0-20\%$  p--Pb collisions at $\sqrt{s_{\rm{NN}}}=5.02$ TeV and in $105 \textless N_{\rm{ch}}\textless 150$ p--p collisions at $\sqrt{s_{\rm{NN}}}=13$ TeV, calculated from  the \hcf{} model and the \hf{} model, respectively.  As a full hybrid model that includes hydrodynamics, quark coalescence and string fragmentation at different transverse momentum regimes, \hcf{} provides a reasonable description of the elliptical flow of identified hadrons over the $p_{\rm T}$ range from 0 to 6 GeV for both p--Pb and p--p collisions. At low $p_{\rm T}$,  hadron productions from \hcf{} are dominated by hydrodynamics, which leads to a clear mass ordering of $v_2(p_{\rm T})$ among different hadron species. At intermediate $p_{\rm T}$, the elliptic flow  calculated with the quark coalescence process shows a grouping behavior for baryons and mesons, where $v_2(p_{\rm T})$ of $\pi$ and K roughly overlap and $v_2(p_{\rm T})$ of P and $\Lambda$ roughly overlap above 2.5 GeV. The experimental data from the CMS and ALICE collaboration also show a similar grouping tendency for baryons and mesons, but need further confirmation with high statistical run data in the near future. In contrast, the \hf{} model largely underestimates  $v_2(p_{\rm T})$ of hadrons for $3 \textless p_{\rm T} \textless 6$ GeV. In this calculation without the quark coalescence contribution, the splitting of elliptic flow among different hadron species arises primarily from their different masses. In particular, regardless of the adjustments of parameters, the \hf{} model cannot simultaneously describe the transverse momentum spectrum and the differential elliptic flow of identified hadrons in either p--Pb or p--p collisions. Note that the quark coalescence procedure in \hcf{} transfers the partonic collectivity at low $p_{\rm T}$ to hadronic collectivity at the intermediate $p_{\rm T}$. Our calculation highlights the importance of the partonic collectivity for describing the differential elliptic flow of light hadrons at the intermediate $p_{\rm T}$ regions in small systems. 
It is noteworthy that the enhanced hadron $v_2$ at $p_{\rm T} \approx 5 - 6$ GeV in the \hcf{} model results from the order of magnitude larger $v_2$ from coalescence compared to those from fragmentation. Although more hadrons are produced by fragmentation than by coalescence at these momenta, the larger $v_2$ from coalescence hadrons contributes significantly to the observed enhancement of $v_2$ at high $p_{\rm T}$ in p--Pb and p--p collisions.

In Fig.~\ref{f:v2nall}, we further study the number of the constituent quark scaling of the elliptic flow in the small systems. The top and bottom panels are the number of constituent quark scaled elliptic flow  $v_2(p_{\rm T})/n$ ($n=2$ for mesons and $n=3$ for baryons) in $0-20\%$  p--Pb collisions at $\sqrt{s_{\rm{NN}}}=5.02$ TeV and in $105 \textless N_{\rm{ch}}\textless 150$ p--p collisions at $\sqrt{s_{\rm{NN}}}=13$ TeV, respectively. 
With the quark coalescence included, the \hcf{} model gives an approximate NCQ scaling behavior in the transverse momentum region $1.2 \textless p_{\rm T}/n \textless 2$  for p--Pb collisions. This result is in agreement with the ALICE measurements. In our calculations, we also observe a slight break of the NCQ scaling, which is due to the contribution from the string fragmentations and resonance decays. In the case of p--p collisions, the break of the NCQ scaling behavior is more noticeable for both the \hcf{} results and the experimental measurements. As shown in Figure~\ref{f:spectraall}, the quark coalescence contributions at intermediate $p_{\rm T}$ are smaller in p--p collisions than in p--Pb collisions, leading to a larger violation of the NCQ scaling in p--p collisions. For comparison, we also perform the \hf{} model calculations without the quark coalescence process. As shown in the two right panels of Fig.~\ref{f:v2nall}, not only is the NCQ scaling behavior at the intermediate transverse momentum region violated, but also the magnitude of $v_2(p_{\rm T})/n$ is significantly underestimated for both p--Pb and p--p collisions.
Compared to the experimental data, the \hf{} model gives a reduced $v_2(p_{\rm T})/n$ for each hadron species by more than 20-50\% in p--Pb and p--p  collisions. This indicates that the quark coalescence process is important in describing the number of the constituent quark scaled elliptic flow  $v_2(p_{\rm T})/n$  for different hadron species in the small systems created at the LHC.

\section{Summary}\label{sec:summary}

In this paper, we study the differential elliptic flow  and its NCQ scaling of identified hadrons in $0-20\%$ p--Pb collisions at $\sqrt{s_{\rm{NN}}}=5.02$ TeV and in $105 \textless N_{\rm{ch}}\textless 150$ p--p collisions at $\sqrt{s_{\rm{NN}}}=13$ TeV, using the \hcf{} model that combines hydrodynamics to produce low $p_{\rm T}$ hadrons, quark coalescence model to produce intermediate $p_{\rm T}$ hadrons, and LBT string fragmentations to produce high $p_{\rm T}$ hadrons. The quark coalescence process consists of the recombination of thermal--thermal, thermal--hard, and hard--hard partons, which are generated by the \vis{} hydrodynamic model and the \lbt{} model, respectively.  

With the combined hadron production procedure, the \hcf{} model quantitatively fits the transverse momentum spectrum of identified hadrons below 6 GeV in both p--Pb and p--p collisions. For the collective flow, the \hcf{} model nicely describes the differential elliptic flow for identified hadrons from 0-6 GeV and the NCQ scaling of the elliptic flow at intermediate $p_{\rm T}$ in $0-20\%$ p--Pb collisions. With the quark coalescence, it also shows a grouping behavior of $v_2(p_{\rm T})$  between mesons and baryons, where $v_2(p_{\rm T})$ of $\pi$ and K roughly overlap and $v_2(p_{\rm T})$ of P and $\Lambda$ roughly overlap above 2.5 GeV. Meanwhile, we also found that the contribution from string fragmentation and resonance decays slightly breaks the NCQ scaling of $v_2$ in high multiplicity p--Pb collisions. For p--p collisions, the \hcf{} model roughly describes the differential elliptic flow of identified hadrons with properly tuned parameters.  However, the quark coalescence contributions at intermediate $p_{\rm T}$ become smaller, leading to a noticeable violation of the NCQ scaling of $v_2$ in p--p collisions, as observed in the experiment. 

Comparison runs of the \hf{} model without the quark coalescence process show that, regardless parameter adjustments, the \hf{} model cannot simultaneously describe the transverse momentum spectrum and the differential elliptic flow of identified hadrons in either p--Pb or p--p collisions, which underestimates the magnitude of $v_2(p_{\rm T})$ at intermediate and high $p_{\rm T}$  regimes and also violates the NCQ scaling behavior of $v_2$ for both p--Pb and p--p collisions. It thus demonstrates the importance of quark coalescence in describing the elliptic flow of different hadrons in the intermediate transverse momentum regime. The calculations in this paper also provide support for the existence of the partonic degrees of freedom and the possible formation of the QGP in the high multiplicity p--Pb collisions and  p--p  collisions at the LHC.

\section{Acknowledgments}
This work is supported by the NSFC under grant Nos. 12247107 and 12075007. W.B.Z. is supported by the National Science Foundation (NSF) under grant numbers ACI-2004571 within the framework of the XSCAPE project of the JETSCAPE collaboration and US DOE under Contract No.~DE-AC02-05CH11231, and within the framework of the Saturated Glue (SURGE) Topical Theory Collaboration.  We also acknowledge the computing resources provided by the SCCAS and Tianhe--1A Super--computing Platform in China and the High--performance Computing Platform of Peking University.

%
\bibliography{pp}
%
\end{document}